# A Multi-Layer Testing Framework for Automated Data Quality Assurance in Cloud-Native ELT Pipelines

Ismail Gargouri(✉) and Hassan Reza

Department of Computer Science, University of North Dakota, Grand Forks, ND, USA

ismail.gargouri@und.edu
hassan.reza@und.edu

**Abstract.** Ensuring data quality in cloud-native Extract–Load–Transform (ELT) pipelines is increasingly challenging due to heterogeneous sources, evolving schemas, and multi-backend execution environments. This paper presents a unified multi-layer testing framework combining orchestration-level validation, declarative dbt tests, LLM-generated semantic tests, and cross-store consistency checking between DuckDB and Snowflake, all orchestrated through Apache Airflow. Controlled anomaly-injection experiments show that the original weak manual-only baseline detected 7 of 16 injected anomalies, whereas both a stronger manual-expanded comparator and the LLM-augmented configuration detected all 16 — a 128.57% relative improvement in detection count. Post-migration cross-store validation confirmed exact agreement across all three curated tables. Of 25 LLM-generated test items, 9 were classified as useful, 4 as redundant, and 12 as executable but low-value. The full workflow completed in 106.577 s across eight instrumented stages, demonstrating that semantic test synthesis can materially strengthen validation coverage while remaining operationally practical.

## 1 Introduction

Modern data-driven systems rely on cloud-native ELT pipelines to support analytics, business intelligence, and machine-learning workloads. As organizations operate across heterogeneous data stores, distributed compute engines, and continuously evolving schemas, ensuring data quality throughout these pipelines has emerged as a foundational software-engineering challenge. Unlike traditional software, data-intensive systems must validate evolving datasets, implicit semantics, cross-source consistency, and transformations executed across multiple compute environments. Failures often propagate silently, compromising downstream analytics and decision making.

Existing solutions — rule-based constraint systems, statistical profiling libraries, and warehouse-integrated frameworks — typically target a single pipeline stage, require substantial manual rule engineering, or lack semantic awareness. The growing reliance on hybrid compute backends such as DuckDB alongside cloud warehouses such as Snowflake introduces a further challenge: correctness must be verified not only within each engine but across them. Prior work offers limited guidance on unified strategies that address structural, semantic, process-level, and backend-level inconsistencies in a single framework.

This paper investigates whether a unified, multi-layer validation framework combining orchestration-level checks, declarative dbt tests, LLM-generated semantic tests, and cross-store consistency verification can improve the reliability of cloud-native ELT pipelines. The framework is implemented within a reproducible Apache Airflow workflow using DuckDB, dbt, Snowflake, and GPT-4.1-mini, and evaluated through controlled anomaly-injection experiments, usefulness auditing, and structured runtime instrumentation. The study is guided by four research questions: (RQ1) can a multi-layer framework improve data-quality issue detection? (RQ2) can LLM-generated dbt tests improve on a weak manual baseline and match stronger human-authored expansion? (RQ3) how reliably does cross-store validation verify DuckDB–Snowflake consistency? (RQ4) what computational overhead does the framework introduce?

This work makes the following key contributions:

- A unified multi-layer framework integrating orchestration validation, declarative dbt testing, LLM-assisted semantic synthesis, cross-store consistency checking, and structured experiment logging in a single cloud-native ELT workflow.
- A reproducible end-to-end architecture combining Apache Airflow, DuckDB, dbt, Snowflake, Amazon S3, and GPT-4.1-mini.
- An empirical comparator-based evaluation showing the weak manual-only baseline detected 7/16 anomalies versus 16/16 for both the manual-expanded and LLM-augmented configurations — a 128.57% relative improvement.
- Evidence that LLM-assisted synthesis matches the detection performance of a stronger human-authored rule set, with its primary practical value lying in automated rule expansion rather than superior raw coverage.
- A structured usefulness audit classifying 25 generated tests into 9 useful, 4 redundant, and 12 executable-but-low-value items.
- Cross-store validation confirming exact row-count and checksum agreement across all three curated tables after DuckDB-to-Snowflake migration.
- Stage-level runtime instrumentation showing a complete research workflow completes in 106.577 s across eight logged stages.

The remainder of this paper is organized as follows. Section 2 reviews prior work. Section 3 presents the methodology and implementation. Section 4 reports results. Section 5 discusses findings, limitations, and future work. Section 6 concludes.

## 2 Related Work

Data-quality validation in cloud-native ELT spans foundational quality models, rule-based systems, pipeline-integrated frameworks, and AI-assisted approaches. Table 1 summarizes twelve representative works across these categories.

**Table 1.** Taxonomy of prior research on data-quality testing in cloud-native ELT pipelines.

| Category | Key Papers | Summary Contribution | Limitations Identified |
|---|---|---|---|
| Foundational models | [4,5] | Multi-dimensional quality definitions; data-cleaning taxonomies. | Assume centralized relational systems; predate cloud-native architectures. |
| Rule-based & statistical | [6,9] | Constraint-based profiling and anomaly detection for batch and ML pipelines. | Manual rule authoring; single-layer; weak orchestrator integration. |
| Cloud-native & pipeline | [1–3,7,8,11] | Schema checking, drift detection, and metadata-driven monitoring in ML pipelines. | No multi-layer testing; no semantic rule expansion; no cross-store checking. |
| AI/ML-assisted | [10,12] | ML-driven anomaly detection and automated quality classification. | No LLM-based dbt test synthesis; no cross-store validation; no ELT integration. |

Wang and Strong [5] established multi-dimensional data quality (intrinsic, contextual, representational, accessibility), while Rahm and Do [4] framed data cleaning as detection, transformation, and verification. Both predate orchestration-driven architectures. Deequ [6] and MLDV [9] automate rule execution but require extensive manual constraint authoring and operate on single pipeline stages. Cloud-native frameworks such as TFX [7,8], and survey works [1–3,11] embed schema checks and drift detection in ML pipelines but provide no multi-layer testing, no declarative dbt support, and no cross-backend consistency verification. AI-assisted works [10,12] explore ML-based anomaly classification but do not study LLM-based declarative constraint synthesis within ELT workflows.

These gaps motivate the present study: no existing work offers a single framework that combines structural rule validation, LLM-generated semantic tests, orchestration-aware execution, and migrated cross-store verification in a reproducible end-to-end ELT setting.

# 3 Methodology and Implementation

## 3.1 System Setup and Architecture

The experimental environment runs on Windows with Docker Desktop (WSL2), Apache Airflow, a Python virtual environment, AWS S3, Snowflake, and an OpenAI API key. The dbt profile targets both DuckDB (local) and Snowflake (cloud). The framework is organized into four modules: airflow/ (DAG and orchestration), dbt_project/ (models, schemas, tests), llm_tests/ (schema extraction, prompt construction, GPT-4.1-mini wrapper, YAML validation, schema merge), and experiments/ (runtime logs, anomaly results, usefulness audits, and research summary). Figure 1 illustrates the conceptual architecture.

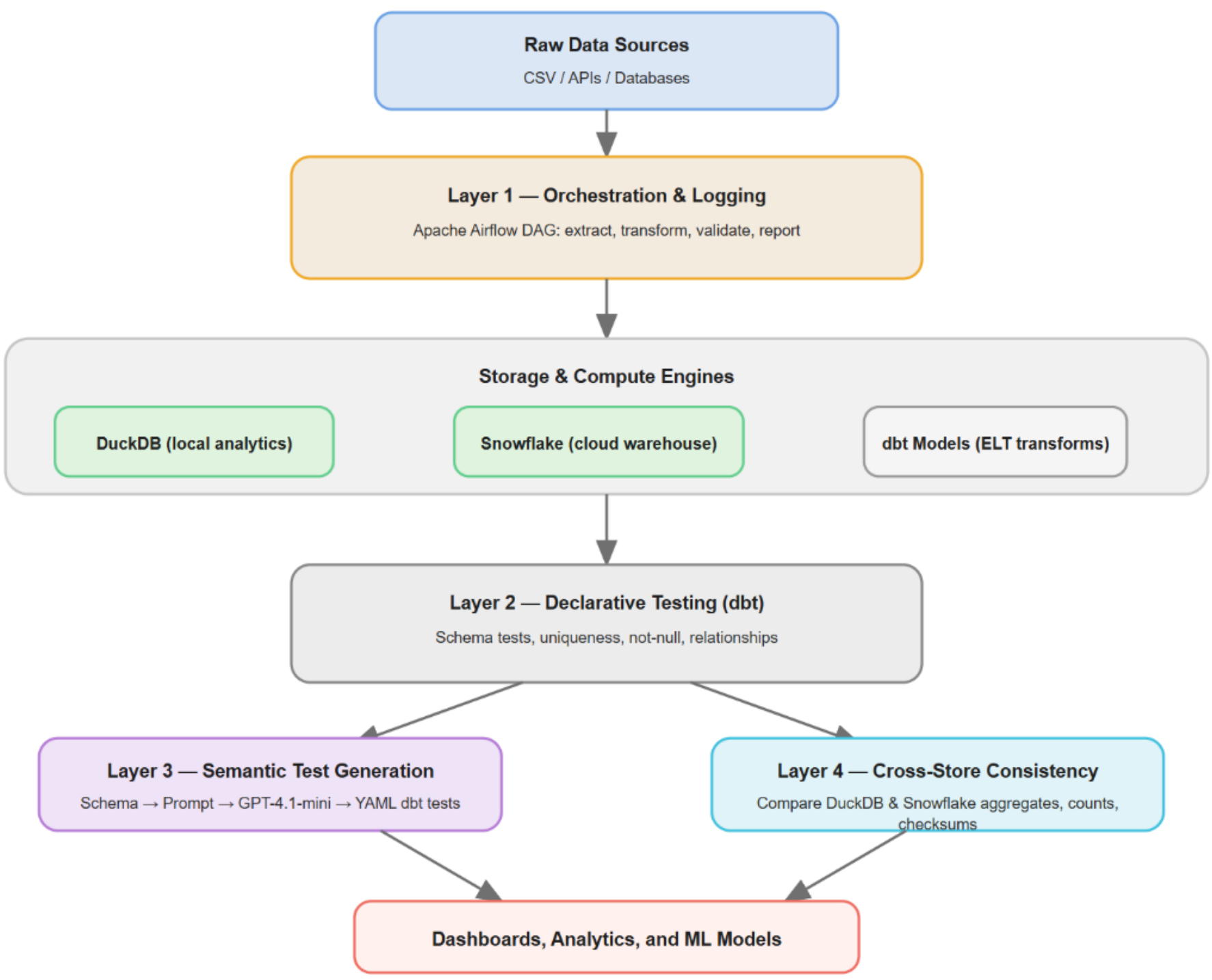

**Fig. 1.** Multi-layer data-quality testing architecture showing the progression from raw data ingestion through orchestration, local transformation, semantic test generation, and cross-store validation.

## 3.2 Experimental Procedure and Orchestration

The full workflow is implemented as an Airflow DAG (football_multi_layer_validation) that enforces a deterministic execution order, summarized in Algorithm 1. Clean-run migration and validation are completed before mutation-based anomaly experiments begin, preventing contamination of evidence.

**Algorithm 1.** Airflow-Orchestrated Multi-Layer Validation Workflow

```
Require: Football API data, dbt project, LLM credentials, Snowflake
credentials
Ensure:  Validated DuckDB outputs, migrated Snowflake tables,
anomaly results, audit
 1: Reset runtime log; restore baseline schema
 2: Ingest raw football data → S3
 3: dbt run --full-refresh (DuckDB)
 4: Extract DuckDB schemas + sample rows; call GPT-4.1-mini → YAML
tests
```

```
 5: Merge generated tests into active dbt schema; execute dbt test
 6: Export curated DuckDB tables → Snowflake (PROD.DUCKDB_MIGRATED)
 7: for each migrated table: compare row counts, checksums, null
summaries, row diffs
 8: for each anomaly batch {A, B, C}: restore snapshot → run
manual-only,
    manual-expanded, manual+LLM; compare detections
 9: Audit 25 generated tests: classify useful / redundant / low-
value / invalid
10: Compile research summary
```

### 3.3 Dataset, Models, and Validation Layers

The evaluation uses the English Premier League 2023–2024 season sourced from a JSON object in S3. Four dbt models are materialized: stg_matches (100 rows), dim_teams (20 rows), fct_matches (100 rows), and fct_training_dataset (100 rows). Only the three curated outputs (dim_teams, fct_matches, fct_training_dataset) are migrated to Snowflake for cross-store validation.

The LLM test generation subsystem targets these three models: it extracts column names, types, and sample rows; constructs structured prompts; submits them to GPT-4.1-mini; validates the YAML output; and merges it into the active dbt schema (preserving a backup of the manual baseline). The anomaly evaluation uses a clean DuckDB snapshot (dbt.clean.duckdb) and three injection batches: Batch A — key integrity (null and duplicate identifiers); Batch B — semantic/domain (null team_name, invalid match_status, missing attributes); Batch C — mixed full-stack (dimension- and fact-table defects with downstream propagation). Each of three conditions (manual-only, manual-expanded, manual+LLM) restores the clean snapshot before execution.

### 3.4 Baseline Configurations and C5 Stability Protocol

**Table 2.** Empirically evaluated validation configurations.

| Configuration | Structural tests | Expanded tests | LLM tests | Primary role in study |
|---|---|---|---|---|
| Manual-only baseline | ✓ | ✗ | ✗ | Weak anomaly-detection reference |
| Manual-expanded comparator | ✓ | ✓ | ✗ | Human-authored comparator |
| Manual + LLM | ✓ | ✗ | ✓ | Automated semantic expansion |
| Full multi-layer pipeline | ✓ | ✓ | ✓ | End-to-end workflow evaluation |

To verify reproducibility, a repeated-trial stability protocol (C5) evaluated the comparator result under two settings: (1) frozen-schema — the merged LLM schema is held fixed and the anomaly comparator rerun five times; (2) fresh-generation — LLM tests are regenerated from scratch before each of five independent comparator runs. Both settings preserve the same three batches and three executable conditions.

## 4 Results and Findings

The framework was evaluated on a controlled experimental setup. Table 3 summarizes all key metrics across the four evaluation dimensions: semantic test generation, anomaly-detection performance, cross-store consistency, and runtime. Figures 2–5 provide visual breakdowns of individual result dimensions.

**Table 3.** Summary of experimental results across the final workflow.

| Metric | Value | Layer |
|---|---|---|
| LLM target models generated successfully | 3 / 3 | LLM semantic |
| Total generated test items | 25 | LLM semantic |
| Useful / redundant / low-value / invalid generated tests | 9 / 4 / 12 / 0 | LLM semantic |
| Manual-only anomaly detection (weak baseline) | 7 / 16 | Baseline |
| Manual-expanded anomaly detection | 16 / 16 | Comparator |
| Manual + LLM anomaly detection | 16 / 16 | dbt + LLM |
| Absolute detection-rate gain (vs. weak baseline) | +56.25 pp | Comparative |
| Relative detection-count improvement (vs. weak baseline) | +128.57% | Comparative |
| Migrated curated tables matching DuckDB ↔ Snowflake | 3 / 3 | Cloud |
| C5 frozen-schema runs: manual+LLM = 16/16 | 5 / 5 | Stability |
| C5 fresh-generation runs: manual+LLM = 16/16 | 5 / 5 | Stability |
| Total instrumented workflow runtime | 106.577 s | Process |

### 4.1 LLM Test Generation Quality

GPT-4.1-mini successfully generated test items for all three target models: dim_teams (3 items: 1 useful, 2 redundant), fct_matches (13 items: 5 useful, 2 redundant, 6 low-value), and fct_training_dataset (9 items: 3 useful, 0 redundant, 6 low-value). No generated test was invalid or non-executable. The most useful tests targeted team_name, match_status, match_date, and downstream completeness in fct_training_dataset. Executability alone is insufficient evidence of value: 12 of 25 tests ran successfully yet contributed no incremental anomaly detection. Figure 2 shows the per-model composition.

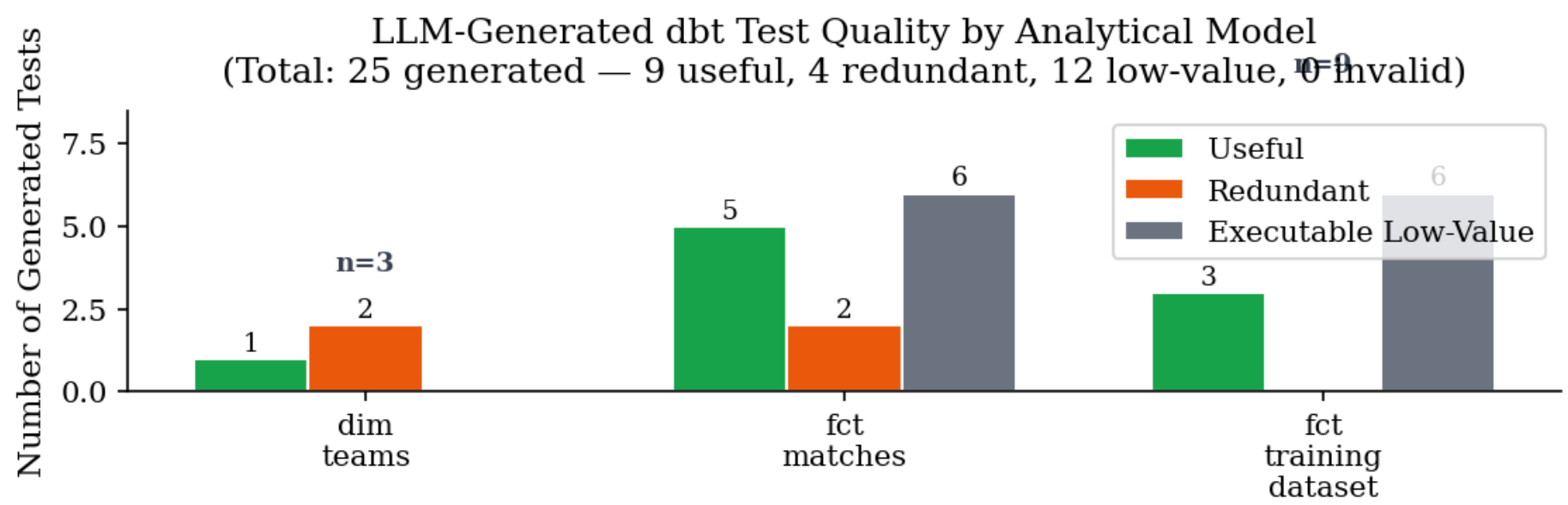

**Fig. 2.** Composition of useful, redundant, and executable low-value LLM-generated dbt tests across the three analytical models.

### 4.2 Cross-Store Validation (DuckDB vs. Snowflake)

All three curated tables — dim_teams (20 rows), fct_matches (100 rows), and fct_training_dataset (100 rows) — were assigned MATCH after migration: row counts were identical, raw checksums matched, null summaries agreed, and zero row-level mismatches were observed after normalization. Figure 3 visualizes the row-count parity and validation status.

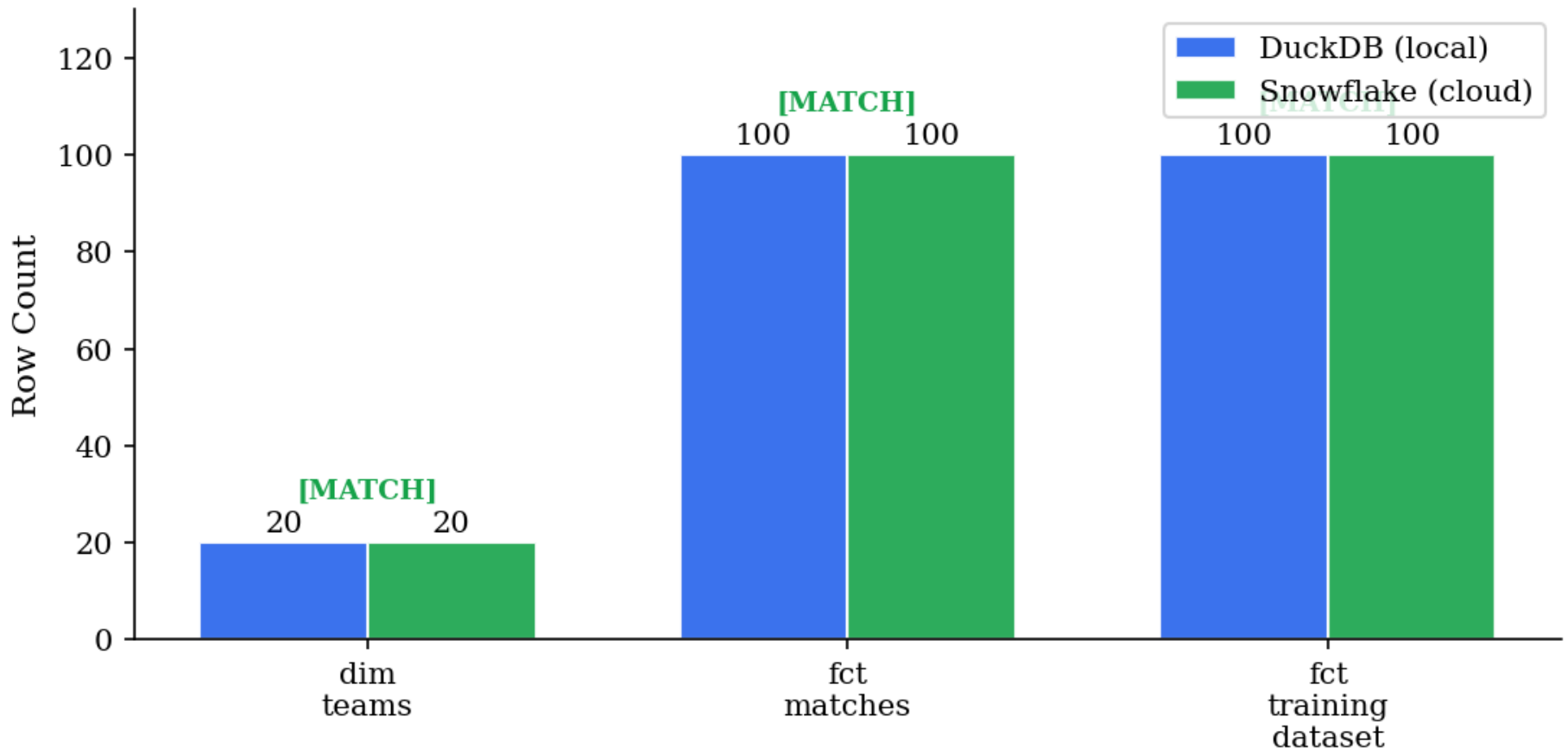

**Fig. 3.** Post-migration cross-store validation results showing row-count parity and MATCH status for all three curated tables.

### 4.3 Anomaly-Detection Performance

The weak manual-only baseline detected 4/4 key-integrity anomalies (Batch A), 0/6 semantic/domain anomalies (Batch B), and 3/6 mixed anomalies (Batch C), totaling 7/16. Both the manual-expanded and manual+LLM configurations detected all 16 across all three batches. The primary gain therefore came from Batch B and Batch C, not from already-covered key-integrity cases. The manual+LLM configuration matched — rather than exceeded — the stronger human-authored comparator, indicating that the LLM layer's value lies in automated rule expansion rather than superior raw coverage. Figure 4 visualizes batch-level detection coverage.

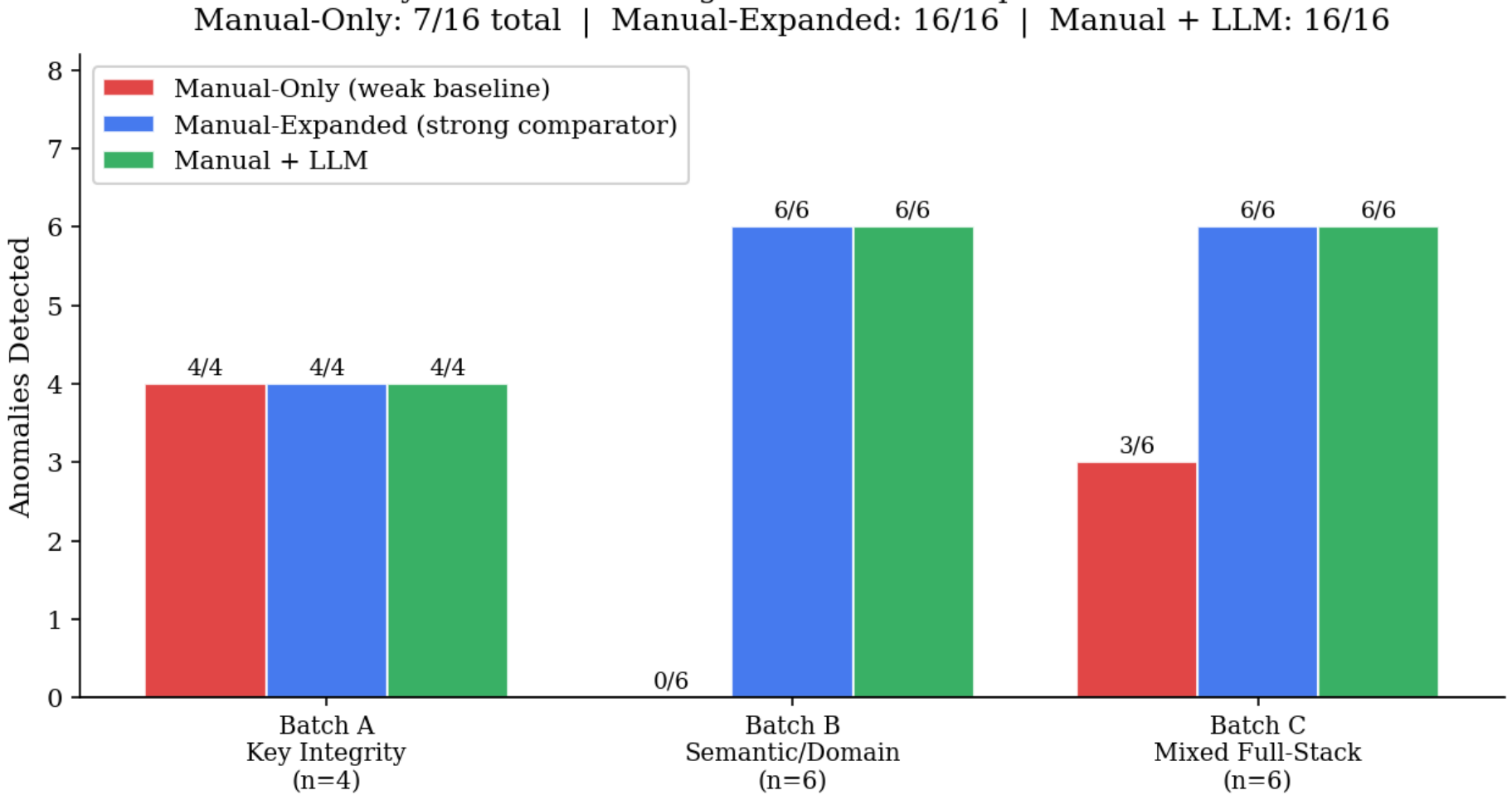

**Fig. 4.** Anomaly-detection coverage of the weak manual-only baseline versus the manual-expanded and manual+LLM configurations across the three experimental batches.

### 4.4 Runtime and C5 Stability

The eight instrumented stages completed in 106.577 s total. The most expensive stages were multi-batch anomaly experimentation (44.095 s), DuckDB-to-Snowflake migration (22.643 s), and migrated-backend validation (14.130 s). LLM test generation required 10.707 s. Clean local dbt run and test stages required 6.997 s and 7.835 s, respectively. Figure 5 shows the per-stage breakdown. Both C5 settings produced fully stable results: in all five frozen-schema reruns and all five fresh-generation trials, the weak baseline remained at 7/16 while both stronger conditions held at 16/16, confirming that the comparator result is a reproducible property of the framework rather than an artifact of a single execution.

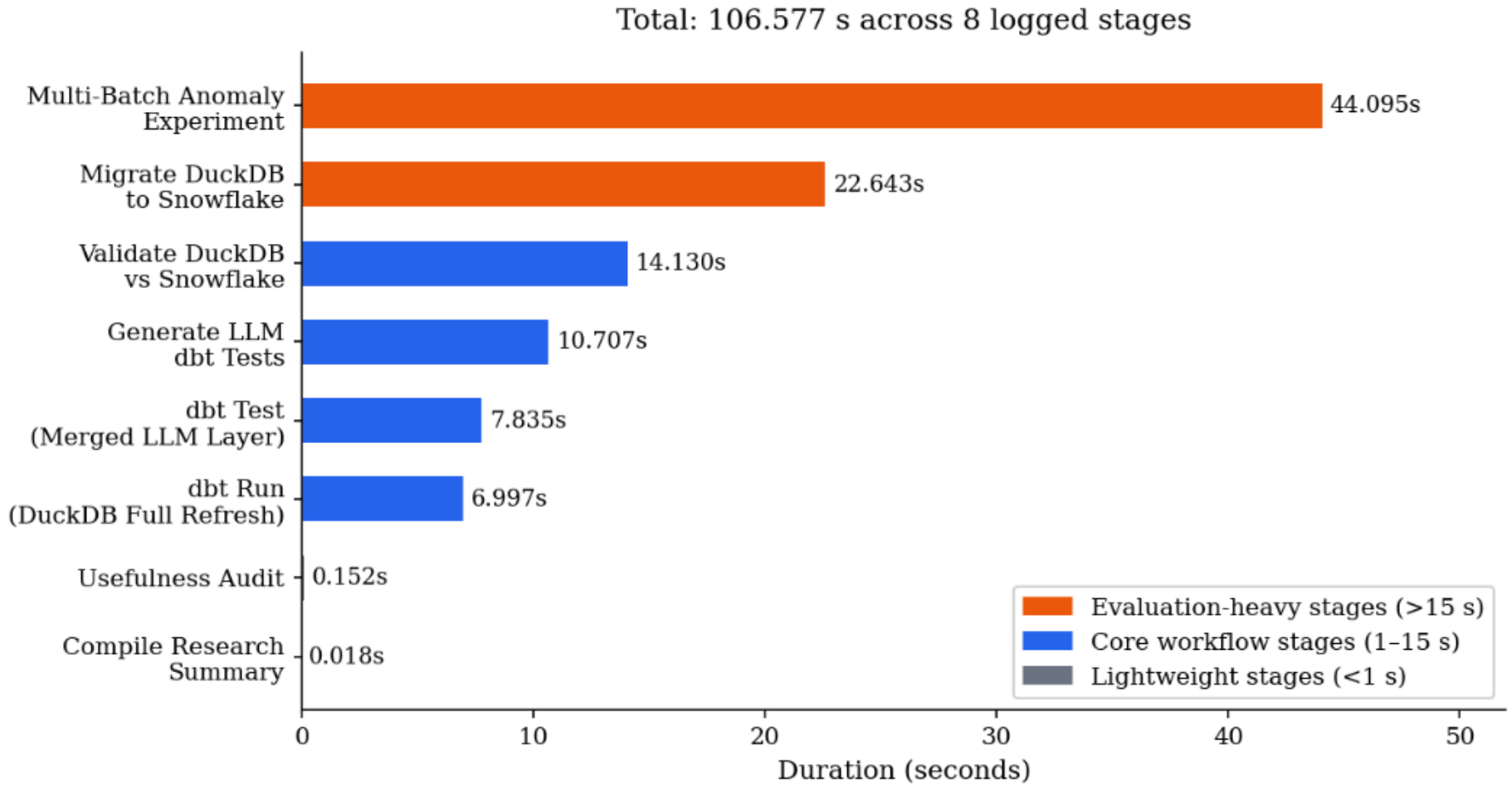

**Fig. 5.** Execution time by instrumented workflow stage, sorted from longest to shortest.

## 5 Discussion

**Multi-Layer Validation Value.** The anomaly-detection results demonstrate that a weak deterministic baseline is insufficient for semantic, completeness-related, and downstream-propagated anomaly classes. Extending that baseline — whether manually or through LLM synthesis — closed detection gaps across Batch B and Batch C entirely, lifting coverage from 43.75% to 100%.

**LLM Contribution.** The LLM-generated layer behaved as a mixed synthesis mechanism: 9 of 25 items added meaningful new coverage, 4 were harmless duplicates, and 12 executed without contributing incremental value. The most useful tests addressed semantically meaningful constraints on categorical attributes and downstream completeness. The comparator result — LLM matching rather than exceeding the human-expanded rule set — suggests that LLM-assisted synthesis is best understood as automated rule expansion that achieves human-comparable depth without equivalent manual effort.

**Cross-Store and Operational Significance.** Exact DuckDB–Snowflake agreement across all three curated tables confirms that backend-level verification can be operationalized as a first-class pipeline stage rather than an implicit assumption. Regarding overhead: the dominant cost drivers are anomaly experimentation, migration, and backend comparison, not the clean local dbt stages. This runtime profile suits scheduled batch-oriented workflows where stronger validation justifies moderate additional execution time.

**Limitations.** The study is limited to a single domain (football analytics), a single LLM configuration (GPT-4.1-mini), and a single backend pairing (DuckDB–Snowflake). The anomaly suite covers three batches and 16 anomalies; richer failure families (temporal drift, cross-table relational corruption, aggregation-level defects) remain unexplored. The usefulness audit is evaluated against the same anomaly batches used to generate tests, introducing a degree of endogeneity; results should be interpreted as indicative of value within this setting rather than as general measures of LLM test quality.

**Future Work.** Planned extensions include: (1) broader anomaly suites covering temporal and aggregation-level failures; (2) multi-LLM comparisons to study the generation quality–executability–runtime trade-off; (3) extension of cross-store validation to additional warehouse and lakehouse backends; and (4) a semi-automated feedback loop for suppressing redundant and low-value generated tests.

## 6 Conclusion

This paper presented a unified multi-layer data-quality testing framework for cloud-native ELT pipelines integrating orchestration-level validation, declarative dbt assertions, LLM-generated semantic tests, migrated cross-store consistency checking, and structured runtime instrumentation. Empirical evaluation on a football analytics workload showed that the manual-only weak baseline detected 7 of 16 injected anomalies, while both the manual-expanded and manual+LLM configurations detected all 16 — a 128.57% relative improvement. The LLM-assisted configuration matched the anomaly-detection performance of a stronger human-authored comparator in all ten C5 stability trials, confirming that its primary value is automated rule expansion rather than superior raw coverage. Post-migration cross-store validation showed exact agreement across all three curated tables, and the complete workflow ran in 106.577 s, demonstrating operational practicality.

Beyond its technical contributions, this work reframes data-quality assurance in cloud-native ELT as a multi-layer testing problem in which structural, semantic, backend, and experimental layers are complementary and collectively more effective than any single layer in isolation. As data systems continue to grow in scale and heterogeneity, such integrated testing strategies will become increasingly important for trustworthy analytics and resilient data infrastructure.

**Acknowledgments.** The authors thank the Department of Computer Science at the University of North Dakota for supporting this research.

**Disclosure of Interests.** The authors have no competing interests to declare that are relevant to the content of this article.

## References

1. Foidl, H., Golendukhina, V., Ramler, R., Felderer, M.: Data pipeline quality: influencing factors, root causes, and processing problem areas. J. Syst. Softw. 210 (2024). https://doi.org/10.1016/j.jss.2023.111946
2. Mbata, A., Sripada, Y., Zhong, M.: A survey of pipeline tools for data engineering. arXiv:2406.08335 (2024)
3. Ehrlinger, L., Rusz, E., Wöß, W.: A survey of data-quality measurement and monitoring tools. Front. Big Data 5 (2022). https://doi.org/10.3389/fdata.2022.850611
4. Rahm, E., Do, H.H.: Data cleaning: problems and current approaches. IEEE Data Eng. Bull. 23(4), 3–13 (2000)
5. Wang, R.Y., Strong, D.M.: Beyond accuracy: what data quality means to data consumers. J. Manag. Inf. Syst. 12(4), 5–33 (1996). https://doi.org/10.1080/07421222.1996.11518099
6. Schelter, S., et al.: Deequ: declarative data validation for large-scale data processing. Proc. VLDB Endow. 11(12), 1781–1794 (2018). https://doi.org/10.14778/3229863.3229876
7. Baylor, D., et al.: TFX: a TensorFlow-based production-scale machine-learning platform. In: Proc. KDD 2017, pp. 1387–1395. ACM (2017). https://doi.org/10.1145/3097983.3098021
8. Caveness, E., et al.: TensorFlow data validation in continuous ML pipelines. In: Proc. SIGMOD 2020, pp. 2793–2796. ACM (2020). https://doi.org/10.1145/3318464.3384707
9. Breck, E., Cabrera, S., Chaudhuri, A., Polyzotis, D.: The ML data validation system. In: Proc. MLSys 2019
10. Felderer, M., et al.: Testing data-intensive software systems. In: Perspectives on Data Science for Software Engineering, pp. 181–200. Springer, Cham (2019). https://doi.org/10.1007/978-0-12-410398-7
11. Ridzuan, N., Idrus, M., Mahdin, H.: A review of data-quality dimensions for big data. IEEE Access 12, 11258–11275 (2024). https://doi.org/10.1109/ACCESS.2024.3353678
12. Azzabi, N., Nafkha, M., Ben Abdallah, R.: A survey on data lake architectures and validation mechanisms. J. Big Data 11 (2024). https://doi.org/10.1186/s40537-024-00900-5
13. Apache Airflow: A platform to programmatically author, schedule, and monitor workflows (2024). https://airflow.apache.org/
14. dbt Labs: dbt documentation: testing, modeling, and transformation framework (2024). https://docs.getdbt.com/
15. Raasveldt, M., Mühleisen, H.: DuckDB: an embeddable analytical database. In: Proc. CIDR 2020
16. Chen, A., et al.: Automated code and constraint generation using large language models. arXiv:2303.05381 (2023)